\documentclass[12pt,preprint]{aastex}
\usepackage{epsfig}

\begin{document}

\title{The GL~569 Multiple System} 

\author{M. SIMON\altaffilmark{1}, C. BENDER\altaffilmark{1}, AND
L. PRATO\altaffilmark{2} }

\altaffiltext{1}{Department of Physics and Astronomy, SUNY,
Stony Brook, NY 11794-3800}

\altaffiltext{2}{Lowell Observatory, 1400 West Mars Hill Rd. Flagstaff, AZ 86001}

\begin{abstract}

We report the results of high spectral and angular resolution infrared
observations of the multiple system GL~569~A and B that were intended
to measure the dynamical masses of the brown dwarf binary believed to
comprise GL~569~B.  Our analysis did not yield this result but,
instead, revealed two surprises.  First, at age $\sim100$ Myr, the
system is younger than had been reported earlier.  Second, our
spectroscopic and photometric results provide support for earlier
indications that GL~569~B is actually a hierarchical brown dwarf
triple rather than a binary. Our results suggest that the three
components of GL~569~B have roughly equal mass, $\sim0.04$~M$_{\odot}$

\end{abstract}

\keywords{stars: binaries: visual --- binaries: spectroscopic ---
stars: low-mass, brown dwarfs --- stars: pre-main sequence ---
techniques: high angular resolution --- techniques: radial velocities}

\section{\sc Introduction}

Accurate and precise measurements of the masses of brown dwarfs are
necessary to validate the theoretical calculations of their evolution
and to determine their location in the HR diagram \citep{bur97,bar03}.  
Astrometric and
spectroscopic observations of binaries provide the means to accomplish
this dynamically.  When the distance to the binary is known, the
astrometric orbit of an angularly resolved binary (hence a ``visual
binary'', VB) determines the total mass, $M_{tot}$.  A double-lined
spectroscopic binary (SB2) provides the mass ratio.  When a binary can
be observed as both a VB and SB2, the component masses can be
determined.  Since the VB also yields the orbital inclination, the VB
and SB2 measurements together provide an independent determination of
the binary's distance.  In the special case of an eclipsing binary,
the orbital inclination is $\sim \pi/2$ and the SB2 observations alone
suffice to determine the component masses.

Although brown dwarf binaries are not rare \citep{burg05}, precise
dynamical masses are known only for the brown dwarf components of the
eclipsing binary in Orion discovered by \citet{stas05}.  Their masses,
radii, and effective temperatures agree reasonably well with the
models, but surprisingly, the effective temperature of the primary is
found to be cooler than that of the secondary.  \citet{zap04},
\citet{bou04}, and \citet{prav05} have reported brown dwarf masses
based on an astrometrically determined $M_{tot}$ and photometry, but
reliable masses determined dynamically are not yet available.
\citet{zap04} obtained angularly resolved spectra of GL~569~Ba and Bb
but their measurements were insufficient to yield dynamical masses.
Because  only two brown dwarf masses are known, and they raise a puzzle
that requires confirmation, it is desirable to enlarge the sample of
dynamically determined masses and to cover as large a range of mass
and age as possible.

GL 569 is a multiple system comprised of GL~569~A, a M2.5V star
\citep{hen90} of mass $M=0.35\pm0.03~M_{\odot}$ \citep{gor03} at the
$HIPPARCOS$ distance $9.81\pm0.16$~pc, and GL~569~B, thought to be a
binary brown dwarf with period $\sim876$~days \citep{zap04}. We began
a study of GL~569~B in 2002 because the astrometric orbit is well
determined and the components, Ba and Bb, are bright enough to be
observed at high spectral resolution in the near-IR.  Thus, GL~569~B
is well suited for the fusion of VB and SB2 techniques to measure the
mass of its components.  This paper presents our results. We describe
our observations and data reduction in \S 2, the results for GL 569 A
and B in \S3 and \S4, respectively, and discuss our results in \S5.
We present a summary and suggestions for future work in \S 6.

\section{\sc Observations and Data Reduction}

All observations were obtained at the W. M. Keck Observatory using
the facility near-infrared spectrometer NIRSPEC \citep{mcl00} and
near-IR camera, NIRC-2.  We used NIRSPEC in its high spectral
resolution mode centered at 1.55 $\mu$m, with and without the adaptive
optics (AO) system \citep{wiz00}. The NIRC-2 observations in the $H$-
and $K$-bands were made using AO. Table~1 provides a log of the
observations.

We used a 2-pixel wide slit for all NIRSPEC observations which
provided spectral resolution $\sim$34,000 with AO and $\sim$31,000
without it.  With AO, the diffraction resolution was $\sim40$~mas.
This enabled us to measure the angularly resolved spectra of GL~569~Ba
and Bb simultaneously with both objects on the appropriately rotated
spectrometer slit.  All observations were taken as a series of
$240-300$~s integrations with the targets shifted in an ABBA pattern
between two positions separated by $\sim$10$\arcsec$ on the slit without
AO, and  $\sim$1$\arcsec$ with AO.
Sets of 10 flats and 10 darks were median filtered and the results,
along with individual AB pairs of target data, were input into the
REDSPEC code for rectification, flat-fielding, and spectral
extraction\footnote{See http://www2.keck.hawaii.edu/inst/nirspec/redspec/index.html}. The resultant spectra for a single object were then
averaged.  We worked with NIRSPEC orders 48 and 49, centered at
$\sim$1.59 and $\sim$1.56$\mu$m, respectively, because these are the
most free of telluric absorption of the 9 orders available at the
grating setting we used \citep[see][Figure~1]{ben05}.  Spectral lamp
line integrations were taken to obtain the dispersion solution for the
AO spectra. For the non-AO observations, the night sky OH lines were
used for this purpose.

We measured radial velocities of the targets with the two-dimensional
correlation algorithm TODCOR \citep{zuc94}, following procedures we
have described previously \citep[e.g.,][]{maz02}. TODCOR requires
template spectra for the primary and secondary components.  When we
observed and reduced NIRSPEC template spectra for the first time
\citep{pra02}, we compared their radial velocities to similar
measurements of the same stars by \citet{duq91} and \citet{lat85} in
order to determine the uncertainty, $\pm$1.0~km~s$^{-1}$, in our
velocities.  More recently, \citet{nid02} have published very precise
radial velocities of 844 FGKM stars, including several whose IR
spectra we use as templates.  We used the \citet{nid02} measurements
to correct the velocities of stars earlier than M4 in our suite of
templates and used the corrected values to bootstrap improvements to
templates of later spectral type. We list the revised values in
Table~2; for the stars in common with \citet{pra02} all the revised
velocities are within 2.2~km~s$^{-1}$ of the values given there, and
most agree to a few tenths of a km~s$^{-1}$.

The NIRC-2 images have a scale of $0\farcs00942$ per pixel and a field
of view of $\sim$$9\farcs6$, sufficient to include both GL~569~A and B
simultaneously.  We took data in the $H$- and $K$-bands in a
five-point dither pattern.  Integration times for each frame varied
from $0.18-1.00$~s.  After flat-fielding the individual images with
the dark-subtracted flat, we expanded the the pixel scale by a factor
of 10 to improve the measurement precision of the relative positions
and photometry for GL~569~A, Ba, and Bb.  For the February, 2005 data,
we used GL~569~A to centroid the images for shifting and adding
multiple exposures and to create an instrumental point-spread function
(PSF) in the $H$- and $K$-bands. GL~569~A was saturated in the images
obtained on December 2004 so we could use these data only for
astrometry and relative photometry of GL~569~Ba and Bb.

\section{\sc Results: GL~569~A}

The proper motion of GL~569~A is $\mu_{\alpha}=275.95$ and
$\mu_{\delta}=-122.12$~mas~yr$^{-1}$ ({\it HIPPARCOS}). Thus, in the
$\sim$20~years since the discovery by \citet{for88} of GL~569~B, the A
component has moved $\sim$$6\arcsec$ on the sky. Our February, 2005
NIRC-2 $H$- and $K$-band images showed the photocenter of GL~569~B
$4\farcs18\pm0\farcs21$ North, and $2\farcs19\pm0\farcs14$ East with
respect to GL~569~A. This offset of B with respect to A lies
$\sim$$1\arcsec$ from that measured by \citet{for88} (Table 3, Figure
1) and confirms their finding that A and B form a common proper motion
system.  Our measurement, taken together with the positions measured
by \citet{mar00} and \citet{lan01}, clearly indicates the orbital
motion of GL~569~B with respect to A (Figure 1).

Using spectroscopy in visible light, \citet{hen90} showed that
GL~569~A's spectral type is M2.5~V.  To measure the radial velocity of
GL~569~A and, in the process, to estimate its spectral type, we
analyzed the order 49 spectra obtained on UT 20 April 2003, UT 14 June,
2003, and UT 25 May 2004. The spectrum of the template star GL~436,
spectral type M2.5~V \citep{pra02}, produced the highest correlation
with GL~569~A's spectrum, in agreement with Henry \& Kirkpatrick's
result.  The difference between the average of our radial velocity
measurements, $-7.7\pm0.6$~km~s$^{-1}$ (Table~4), and the
center-of-mass velocity of the Ba$-$Bb pair (\S 4.3) is
$0.8\pm0.9$~km~s$^{-1}$.  The velocity difference arising in the
orbital motion of B with respect to A is expected to be small (\S 5)
and to lie below the significance limit of our measurements.

The apparent magnitudes of GL~569~A at $V$ and $H$ are
$10.200\pm0.004$~mag and $5.99\pm0.02$~mag, respectively ({\it
HIPPARCOS, 2MASS}), corresponding to absolute magnitudes
$M_V=10.24$~mag and $M_H=6.03$~mag at distance 9.81~pc.  Using near-IR
surface gravity indicators and model stellar spectra, \citet{gor03}
estimated that GL~569~A's mass is $0.35\pm0.03$~M$_{\odot}$.  Figure 2
shows $M_V$ and $M_H$ vs mass diagrams using these values.
Surprisingly, these mass-luminosity diagrams indicate an age of
20$-$40 Myr for GL~569~A, much younger than the several hundred Myr
estimates of the ages of the brown dwarfs Ba and Bb \citep{lan01,
zap04}. It is possible, however, that the theoretical isochrones 
underestimate the age of a star in this range of mass and age 
(I. Baraffe, priv. comm.).
To obtain an independent estimate of the age, we followed
K.~Luhman's suggestion (priv. comm. 2005) and placed GL~569~A on a
(V$-$K) color vs. absolute $K$ magnitude diagram (CMD) for the
Pleiades cluster, thought to be 100$-$125 Myr old \citep{mey93,sta98}.
Figure~3 shows that GL~569~A lies on the Pleiades main-sequence,
indicating that it is about the same age.  
GL~569~A could be older than it appears in Figures~2 or 3 if it were fainter
and its apparent magnitude represented that of two stars in a binary.
However, its spectrum in both the visible and IR is that of an M 2.5 with
no evidence of either a companion, or variability of its velocity on
the three occasions that we measured it in the IR.  Its H and K-band
images also show no evidence of a companion. We conclude that 1) GL 569 A 
is a single star, 2) its age is 100-125 Myr,  and 3) GL 569 A seems 
to be younger than the age previously estimated for its brown dwarf companions.

\section{\sc Results: GL~569~B}

\subsection{Photometry, Astrometry, and Orbital Parameters of
GL~569~B as a Visual Binary} 

Figure 4 shows the $H$- and $K$-band composite images of GL~569~Ba-Bb
from the UT 25 February 2005 NIRC-2 observations.  We modeled the
brown dwarf binary as the sum of two scaled PSFs; the single star PSF
was provided by the simultaneously obtained image of GL~569~A, which
had FWHM = 41~mas at $H$ and 53~mas at $K$.  The AO correction
achieved a Strehl ratio of $\sim$0.3 at $H$ and $\sim$0.4 at $K$.  We
determined the positions and amplitudes of Ba-Bb relative to each
other, and to GL~569~A, by least squares fitting.  We used the 2MASS
magnitudes of GL~569~A at $H$ and $K$, $5.99\pm0.02$~mag and
$5.77\pm0.02$~mag, respectively, to determine the magnitudes of Ba and
Bb. The astrometric and photometric results are given in Table 5.
Because we could not use GL~569~A as a PSF (\S 3) in the December,
2004 data, we formed a composite Ba-Bb PSF by cross-correlating the
images and  shifting them for greatest correlation.  We measured the
position of Bb relative to Ba by modeling the PSF as an elliptical
Gaussian and fitting to the composite image.  The astrometric results
and Bb/Ba flux ratios from the December, 2004 observations are
included in Table 5.  Ba is $0.53\pm0.08$ magnitudes brighter than Bb
at $K$, consistent with the measurement of \citet{lan01}, and
$0.61\pm0.07$ magnitudes brighter at $H$.

Observations of a VB
yield $P, a(''), e$, $i$, $T_o$, $\omega$, and $\Omega$ -- the orbital
period, semi-major axis in arcseconds, eccentricity, inclination, time
of periastron passage, longitude of periastron, and longitude of the
ascending node, respectively\footnote{$\omega$ and $\Omega$ determined
for a VB are degenerate by 180$^\circ$; the ambiguity is removed by
observation of the system as a spectroscopic binary.}. When the
distance to the binary is known, as it is here, the semi-major axis,
in physical units, and period determine the total mass of the binary
via Kepler's Third Law.  Figure 5 shows the apparent orbit derived
from the astrometry reported by \citet{lan01}, \citet{zap04}, and
Table 5.  We calculated the orbital parameters using procedures
written by \citet{sch05} and described in \citet{sch03}.  The new
values (Table 6) are in excellent agreement with those published by
\citet{zap04}. Our observations extend the time over which the orbital
motion of the GL~569~B binary has been monitored from 0.85 to 2.3
orbital periods; as expected, their main effect is to improve the
precision of $P$ and $T_o$. The total mass of Ba$-$Bb that follows from
these values is $M_{tot}=0.125\pm0.007$~M$_{\odot}$. This value is
consistent with that determined by \citet{zap04} but the uncertainty
is larger because our value includes the uncertainty of the {\it
HIPPARCOS} distance measurement and theirs does not.

\subsection{Velocities of the GL~569~B Spectroscopic Binary}

Non-AO NIRSPEC spectra of GL~569~B do not angularly resolve Ba and Bb.
A TODCOR analysis of blended spectra of GL~569~B using our suite of
templates (Table~2) showed that the template spectrum of GL~644~C (M7)
provided high correlation. The analysis also showed that precise
measurement of the component velocities would be difficult because the
radial velocity difference between Ba and Bb is only a few
km~s$^{-1}$, the spectra of the components were similar and complex,
and the spectral lines of Ba appeared to be significantly broadened.  
We expected that we
would obtain the highest velocity precision by using angularly
resolved spectra of Ba and Bb themselves as templates.  To this end,
we used NIRSPEC with AO to measure their angularly resolved spectra on
UT 20 April 2003 and UT 24 May 2004.  The small angular separation of Ba and
Bb required care in the extraction of their spectra to minimize contamination.
Figure~6 illustrates our procedure.  The lower panel shows a sample profile 
of the two-dimensional spectrum in the
cross-dispersion direction from UT 24 May 2004, obtained by median
filtering a REDSPEC rectified image along the dispersion direction; the
AO plate scale is $0\farcs0182$ pixel$^{-1}$. There is significant
blending in the region between Ba and Bb.  We modelled the profile as the
sum of two Gaussians of equal width with separation equal to that of Ba and Bb.
The upper panel of Figure~6 shows the extraction widths, denoted by the 
vertical lines and hashed regions, chosen to avoid the central
region where the profiles are blended. The FWHM
of the Gaussians in Figure~6 is about twice the diffraction limit at 
$1.6\mu$m.  The quality of the seeing and the AO correction varied during
the May 2004 observations; the PSF in the profile shown was especially
broad.  We varied the extraction widths to suit each observation.  Figure~6
shows that even in this case the selected regions suffer little
contamination.  We believe therefore that our procedure produced 
individual target spectra and hence templates useful for analysis
of the blended spectra.

To determine the radial velocity of Ba in orders 48 and 49 on 
UT 20 April 2003 and UT 24 may 2004,  we cross-correlated spectra
of Ba with our GL~644~C template, and found the highest
correlation for $v_{rot}sini = 25$~km~s$^{-1}$, if we interpret the line
broadening as attributable to rotation. Similarly, for Bb, using the GL~644~C
template, the highest correlation was obtained with $v_{rot}sini =
10$~km~s$^{-1}$. The measured radial velocities of Ba and Bb in the
two orders on the two dates differed by the measurement error.  To use
the individual Ba and Bb spectra as templates to analyze the blended,
non-AO B component spectra, we averaged the measured velocities in
orders 48 and 49; these values appear in Table~7.

Figure~7 shows, in the heliocentric reference frame, the
April, 2003 and May, 2004 AO resolved spectra of GL~569~Ba and Bb in
orders 48 and 49 and, for comparison, a non-AO spectrum of GL~569~B
measured on UT 22 February 2005.  The spectral range displayed for
order 48 is less than that for order 49 because we did not analyze the
$\sim$$0.007\mathrm{\,\mu{}m}$ wide section shortward of
$1.585\mathrm{\,\mu{}m}$ that suffers from terrestrial $\mathrm{CO_2}$
absorption \citep[see][Figure~1]{ben05}.  It is evident that the
lines in the  spectrum of Ba are more broadened than those of Bb.  
Except for the
effects of broadening, the spectral features in Ba and Bb appear
identical suggesting that the spectral types must be very similar.
These results are consistent with the finding of \citet{zap04} that,
in their high-resolution, angularly resolved $J-$band spectra, also
obtained using NIRSPEC, Ba and Bb appeared as spectral type M8.5 and
M9, with significant broadening characterized by $v_{rot}sini \sim
37~\mathrm{km~s^{-1}}$ and $30~\mathrm{km~s^{-1}}$, respectively.

We analyzed the non-AO spectra with TODCOR twice using the two
different AO-resolved templates for Ba and Bb from the two different NIRSPEC 
with AO runs.  We list the radial velocities in Table~7.  The OH emission
lines in order 48 are weaker than those in order 49; on UT 22
December 2002 and UT 14 June 2003, the integration times were not long
enough to expose the OH lines for a reliable dispersion solution and
we do not list order 48 velocities for those dates.  We considered the
difference in velocities derived on a given date using the two sets of
AO-resolved templates as an empirical measure of the error.  Thus, for
each order, we considered the differences

$$\Delta_{Ba,i} = V_{Ba,i}(4/03) - V_{Ba,i}(5/04) $$

\noindent in which the $V_{Ba,i}$'s are the radial velocities of Ba
measured on the $i$th occasion using the AO templates obtained in
April, 2003 and May, 2004 (Table~7).  The same procedure was applied
to the velocities of Bb.  The distribution of the 28 differences is
centered at $0.1\pm1.7$~km~s$^{-1}$, indicating that there is no
systematic velocity offset in either set of templates.  The standard
deviations for the $\Delta_{Ba,i}$ and $\Delta_{Bb,i}$ are 1.8 and
1.4~km~s$^{-1}$, respectively; these values are listed in Table~7 as
the velocity uncertainties of the Ba and Bb components.  We attribute
the larger uncertainty associated with the measurements for Ba to the
greater velocity width of its spectral lines.

\subsection{Dynamical Masses of the Components}

Velocity measurements of a double-lined spectroscopic binary (SB2)
yield $P, e, T_o, \omega$, $asini$, in physical units, and the
velocity semi-amplitudes of the primary and secondary, $K_{Ba}$ and
$K_{Bb}$. These are related to the mass ratio, $q=M_{Bb}/M_{Ba}$, $a$,
and $a_{Ba}$ and $a_{Bb}$, the semi-major axes of the primary and
secondary by $q=K_{Ba}/K_{Bb}$ and
$asini=(a_{Ba}+a_{Bb})sini$. $a_{Ba}sini$ is given by

$$ a_{Ba}sini=0.01375K_{Ba}P(1-e^2)^{1/2} $$

\noindent where $a_{Ba}$ is in gigameters, $K_{Ba}$ in km~s$^{-1}$, and
$P$ in days.  The same result holds for $a_{Bb}$ in terms of $K_{Bb}$
\citep{hei78}.

Here, $P, e, i, a(''),$ and $T_o$ are already known to high precision
from the astrometric observations of GL~569~B.  In combination with
the $HIPPARCOS$ distance of GL~569~A, $a$ is known in physical units
(Table 6).  Our analysis of GL~569~B as an SB2 began by calculating
the difference, in the $\chi^2$ sense, of possible velocity solutions
to the data, deriving $K_{Ba}$, $K_{Bb}$ and $\gamma$, the
center-of-mass velocity, for the minimum value of $\chi^2$.  We did
this twice, the first time applying the templates for Ba and Bb
measured in April, 2003, and the second using the May, 2004 templates.
The minimum reduced $\chi^2$ is $\sim$1.1 for both solutions. The
results for the velocity semi-amplitude of Bb and the center-of-mass
velocity are in excellent agreement with each other, but the results
for the velocity semi-amplitude of Ba are not.  Using the April, 2003
templates,

$$ K_{Ba}=2.90\pm 0.50, K_{Bb}=5.34\pm 0.50,
\gamma=-8.47\pm0.30~\mathrm{km~s^{-1}}$$

\noindent and using the May, 2004 templates,

$$ K_{Ba}=4.30\pm 0.70, K_{Bb}=5.30\pm 0.50,
\gamma=-8.50\pm0.30~\mathrm{km~s^{-1}}.$$

We attribute the large difference of the two $K_{Ba}$ values to the
large velocity width of Ba's spectrum, as also manifested (\S 4.2) in
the scatter of the $\Delta_{Ba,i}$, larger than that of the
$\Delta_{Bb,i}$.  Because $K_{Bb}=5.3\pm0.5~\mathrm{km~s^{-1}}$ is
well-determined, we analyzed GL~569~B as single-lined system on the
basis of the Bb data.  Figure~8 shows the velocity $vs$ phase data for
the April, 2003 and May, 2004 templates.  The velocities of the
secondary are fitted with $K_{Bb}$ and $\gamma$ as given above.  Using
the April, 2003 data, the mass function $f(M)$ of the secondary is,

$$f(M) =  {(M_{Ba}sin~i)^3 \over  (M_{Ba}+M_{Bb})^2}$$

$$ =1.036\times 10^{-7} K_{Bb}^3 P (1-e^2)^{3\over 2} ~M_{\odot} $$

\noindent with $K$ in km~s$^{-1}$ and $P$ in days \citep[adapted
from][]{hei78}.  Using $K_{Bb}=5.3\pm0.5$~km~s$^{-1}$ and values for
$P, e, i$ and $M_{Ba}+M_{Bb}$ from Table 6, we derive that
$M_{Ba}=0.105\pm0.011$~M$_{\odot}$.  By subtraction from the total
mass, $M_{Bb}=0.020\pm0.013$~M$_{\odot}$.

The value of $M_{Ba}$ cannot be taken at face value given the
similarity of the Ba and Bb spectra (Figure~7) and their flux ratio
$\sim0.6$ (Table~5).  A promising explanation is that Ba itself is a
close, angularly unresolved binary, and that GL~569~B consists of
three brown dwarfs, two at Ba, one at Bb as previously suggested by
\citet{mar00} and \citet{ken01}. The masses of the Ba brown
dwarfs are probably $\sim$$0.045-0.050$~M$_{\odot}$ each, and that of
Bb, $\sim$$0.020-0.040$~M$_{\odot}$.  This explanation would account
for the observed facts: that Ba is about twice as bright as Bb at H,
that the total mass is $\sim$0.12~M$_{\odot}$, and that the spectra of
Ba and Bb appear identical except for the greater line width of Ba.
The greater line width of Ba could
be attributable either to rotational broadening, which could disguise
the presence of an additional component, or some combination of line
broadening and Ba orbital motion.  If the candidate 
binary possesses an orbital inclination distinct from and larger than
that of the Ba$-$Bb system, the confusion between broadened lines and
velocity shifts would be accentuated, particularly for objects of the similar 
spectral type.  More AO-resolved, high resolution spectra of Ba are required to
distinguish among these possibilities.

Figure~9 shows GL~569~Bb on a mass-luminosity (at $H$) diagram; if its
mass at the $1\sigma$ upper bound is 0.033~M$_{\odot}$, the isochrones
indicate a $\sim$100~MY age, consistent with that of GL~569~A. We do
not know the individual $H-$magnitudes or masses of the brown dwarfs
proposed to comprise GL~569Ba.  If their masses are
$\sim$0.045~M$_{\odot}$ with M$_H$ a few tenths brighter than 11~mag,
the isochrones indicate an age $>100$~MY.  The uncertainties are
suffficiently large that it is unclear whether these mass estimates
are in conflict.

The derived center-of-mass velocity of Ba$-$Bb is well determined at
$\gamma=-8.50\pm0.3$~km~s$^{-1}$. It differs, however, by 3~km
s$^{-1}$ from the value reported by \citet{zap04},
$-11.5\pm0.5$~km~s$^{-1}$. Unlike our measurement, which is referenced
to the stellar radial velocities of the spectral type templates,
\citet{zap04} measured the Doppler shift of the K~I doublet in their
$J$-band spectra using the observed and laboratory wavelengths.  We
believe that our value is representative of the actual value because
it is based on all the lines in the spectra.  Our measurement thus
avoids the complications of measuring a Doppler shift of the pressure
broadened and nearly saturated lines of the K~I doublet.  
Moreover, for M spectral
types, contamination by Mg~I at 1.24369~$\mu$m and particularly by
Cr~I at 1.25253~$\mu$m for later types can shift the line centers of
the K~I doublet at 1.24356~$\mu$m and 1.25255~$\mu$m.  The fact that
our measured center-of-mass velocity differs from the radial velocity
of GL~569~A by only $0.8\pm0.9$~km~s$^{-1}$ also suggests that it is
close to the actual value.  The $\sim$3~km~s$^{-1}$ difference between
our value and that of \citet{zap04} affects only their absolute radial
velocities.  The differences in the radial velocities of Ba and Bb
measured by \citet{zap04} (their Table~4 and Figure 8) agree well with
the differences derived from our values in Table~7, as shown in our
Figure~10.

\section{\sc Discussion}
 
\subsection{The GL~569 Quadruple?}

Our work suggests that GL~569 may be a quadruple consisting of an
M2.5~V star (GL~569~A) accompanied by three nearly equal mass brown
dwarfs and distributed as follows:

\noindent 1) The wide binary, GL~569~A $-$ GL~569~B, at
separation $\sim$5$\arcsec$ ($\sim$50~AU) with total mass
($0.35\pm0.03+0.125\pm0.007)=0.48\pm0.03$~M$_{\odot}$. If it is oriented
face-on, its orbital period is $\sim$500~yrs.

\noindent 2) The binary GL~569~Ba $-$ GL~569~Bb which
has semi-major axis 90~mas (0.89~AU), period
864~days, and total mass $0.125\pm0.007$~M$_{\odot}$.  Bb is a single
brown dwarf with mass $0.020 \pm 0.013$~M$_{\odot}$.

\noindent 3) Ba is a candidate unresolved inner brown dwarf binary with 
total mass $0.105\pm0.011$~M$_{\odot}$.

Designating the two components of Ba, the inner binary, as Baa and
Bab, we consider the limits that can be placed on their orbit
empirically by our spectroscopic and imaging observations, and
theoretically by the requirement that Ba and Bb form a stable
triple. By simulating the inner binary as composed of two equal
brightness brown dwarfs of late M spectral type, rotationally
broadened to 10~km~s$^{-1}$, we found that we could measure their
individual velocities reliably to $\sim$5~km~s$^{-1}$.  We measured no
significant velocity differences in the two AO spectra of Ba.  We
conclude that the observed velocity difference of the two components
was $\le 5$~km~s$^{-1}$ at the times of the observations if the
components had $\sim 10$ km~s$^{-1}$ rotational broadening.  With
$M_{tot}$ = 0.105~M$_{\odot}$, this implies that the semi-major axis of
the inner binary satisfies $a_{inner} \ga 3.8\,\mathrm{sin^2} i_{inner}$~AU,
where $i_{inner}$ is the inclination of the inner binary.

The FWHM of the H-band PSF was 41~mas in the images obtained on UT 25
February 2005.  We analyzed images of model binaries with equal
brightness components convolved with the PSF and the same signal to
noise as the H-band image in Figure~4.  We were able to identify the
two components at separations as small as $\sim 10$ mas.  The actual
situation is complicated by the presence of the third component, Bb,
at apparent separation $\sim$~85~mas (Table 5), $2\times$FWHM of the
PSF.  It is apparent in Figure~4 that the broad wings of the PSF
extend from Bb to Ba. We estimate that Baa and Bab would have to lie
at apparent separation $\ga$~20~mas for their presence to be
recognizable in our images.  This limit, combined with that from the
spectroscopic observations, restricts $i_{inner}$ to $\la
\pm13.6^{\circ}$.

Theoretical study of the stability of triples has an extensive
literature and continues to be an interesting area of research as new
environments are identified (e.g., the formation of planets in stellar
binaries, Holman \& Weigert 1999) and the meaning of ``stability'' is
sharpened (e.g., Ford et al 2000).  \citet{egg95} presented an
empirical stability criterion in terms of $Y^{min}$, the ratio of the
periastron distance of the outer binary to the apastron distance of
the inner binary,

$$ Y^{min}= {a_{outer}(1 - e_{outer}) \over a_{inner}(1+e_{inner})} $$

\noindent When values of $Y^{min}$ larger than a critical value
 $Y^{min}_{0}$ are realized, the triple is stable.  The criterion is
 applicable to triples with various inclinations, eccentricities, and
 relative phases of the inner and outer binaries. For a triple with
 equal mass components \citep[see][eqn. 2]{egg95}, $Y^{min}_{0} \sim
 5$.  With $a_{outer}=0.90$~AU and $e_{outer}=0.31$ (Table 6),
 $a_{inner}(1+e_{inner})=0.13$~AU; $a_{inner}= 0.13$ AU if
 $e_{inner}=0$, and 0.07~AU if $e_{inner} = 0.9$.  In the latter case,
 the apastron of the inner binary is 0.13~AU.  We conclude that the
 largest value of $a_{inner}$ for stability, 0.13~AU, is consistent
 with our observational limits if $i_{inner}$ lies in the range
 $\pm11^{\circ}$.

\subsection{The Lithium Test}

The Li spectral line at 6708~\AA{} is a sensitive mass and age
indicator because brown dwarfs with mass $> 0.06$~M$_{\odot}$ destroy
their Li within a few hundred Myr \citep{mag93,cha00}.  The masses we
have estimated for Baa, Bab, and Bb are at this boundary.  In
spectroscopic observations that did not resolve the GL~569~B system,
\citet{mag93} did not detect the Li line, suggesting that the brown
dwarf masses are greater than the values we infer. Detection of the Li
feature in GL~569~B may have been frustrated by the large line widths
revealed by our analysis of the IR spectra and the possible triple
nature of the system.

\section{Summary and Suggestions for Future Work} 

There are two main results of our work:

1) The previous suggestions that Ba could be a binary brown dwarf were based
on the fact that Ba is $\sim 1.6$ times brighter than Bb in the near IR and
that the near IR colors of the two are similar (Mart\'{\i}n et al. 2000; 
Kenworthy et al. 2001).
Our work confirms these observations and strengthens the case for the binarity
of Ba in three respects: {\it i)} the finding that $M_{Ba} =0.105\pm 0.011
~M_{\odot}$ and Ba cannot be a star of this mass, {\it ii)}  
the similar spectra of Ba and Bb indicate that Ba and Bb contain similar 
brown dwarfs, and {\it iii)}  the greater width  of the  lines of Ba, 
$\sim 25$ km s$^{-1}$, than that of Bb,  $\sim 10$ km s$^{-1}$, has a 
natural explanation if Ba harbors two brown dwarfs.

These arguments and the total mass of the system $M_{tot}=0.125\pm 0.007~
M_{\odot}$ also suggest that the masses of the three components Baa, Bab,
and Bb, are approximately equal and in the range 0.04 to 0.05 $M_{\odot}$.

2) The location of GL 569 A with respect to theoretical isochrones and in
the Pleiades CMD, and the location of Bb with respect to theoretical
isochrones (and also of Baa and Bab if they have approximately equal mass)
indicate that the age the GL 569 system is 100-125 Myr.

The GL~569 quadruple is well suited to improve our understanding of
objects at the low mass end of the IMF.  We suggest
the following observations to confirm and advance beyond the results
we have presented.

\parindent=0.0in

A) Detection of the Li line in the spectra of GL~569~Ba and Bb both
would confirm the mass of GL~569~Bb and that Ba must be composed of two
brown dwarfs. The spectroscopic measurement should resolve Ba and Bb
to avoid the complications of interpreting the complex spectrum of
three unresolved brown dwarfs. This would require AO-assisted
spectroscopic observations in the visible, a capability not yet
available.

B) Additional high angular resolution IR spectroscopy of Bb will
improve the precision of the velocity semi-amplitude $K_{Bb}$, and
hence of $M_{Ba}$ and $M_{Bb}$.

C) High angular resolution IR spectroscopy of Ba may distinguish the
two brown dwarfs Baa and Bab, and hence determine their parameters as
a double-lined spectroscopic binary.

D) At separation $\sim$7~mas, the inner binary is resolvable by the
present generation of IR interferometers and is not far below their
present sensitivity limits. The VB parameters of the inner binary,
combined with the SB2 parameters would determine the individual masses
$M_{Baa}$ and $M_{Bab}$ and reveal whether the orbits in the
hierarchical triple are coplanar.

Of the observations suggested only the AO-resolved IR spectroscopic
observations B and C are possible now.  We plan to continue with these and
look forward to the technical advances that will enable the
interferometric and visible light spectroscopic studies.


\acknowledgments We acknowledge with pleasure I. Baraffe, L. Close, 
J. Lissauer, K. Luhman, and J. Stauffer for their helpful advice and thank
K. Luhman for providing us with the Pleiades data used in Figure~3.
We thank G. Schaefer for use of her computer programs for the analysis
of astrometric orbits and S. Zucker and T. Mazeh for the use of their
TODCOR program for the measurement of velocities in blended
spectra. We are grateful to A. Tokunaga for collaborating on initial
observations at the Subaru telescope. We thank the Keck Observatory
OAs and staff for their expert assistance.  This research was supported in
part by NSF Grant 02-05427 to MS and NSF Grant 04-44017 to LP.  LP
also acknowledges support from Keck PI Data Analysis funds,
administered by JPL for Keck time allocated through NASA.  The authors
benefitted from time allocations through the NOAO-administered NSF
TSIP program.  Data presented herein were obtained at the W. M. Keck
Observatory, which is operated as a scientific partnership between the
California Institute of Technology, the University of California, and
NASA. The Observatory was made possible by the generous financial
support of the W. M. Keck Foundation.  The authors wish to extend
special thanks to those of Hawaiian ancestry on whose sacred mountain
we are privileged to be guests.  Our research used the SIMBAD
database, operated at CDS, Strasbourg, France and data products of
2MASS, which is a joint project of the Univ. of Massachusetts and IPAC
at the California Institute of Technology, funded by NASA and NSF.


\clearpage

\begin{deluxetable}{lrlll}
\tablecaption{Log of Observations}
\tablewidth{0pt}
\tablecolumns{3}
\tablehead{
  \colhead{UT Date} &
  \colhead{MJD} &
  \colhead{Instrument} &
  \colhead{Mode} &
  \colhead{Targets}
  }
\startdata
2002 Jul 18 & 52473.3 & NIRSPEC & non-AO & GL569B \\
2002 Dec 22  & 52630.7 & NIRSPEC & non-AO & GL569B \\
2003 Mar 25  & 52723.5 & NIRSPEC & non-AO & GL569B \\
2003 Apr 20  & 52749.4 & NIRSPEC & AO     & GL569A,GL569Bab \\
2003 Jun 14 & 52804.2 & NIRSPEC & non-AO & GL569A,GL569B \\
2004 Jan 26  & 53030.6 & NIRSPEC & non-AO & GL569B \\
2004 Apr 29  & 53124.4 & NIRSPEC & non-AO & GL569B \\
2004 May 24  & 53149.3 & NIRSPEC & AO     & GL569Bab \\
2004 May 25  & 53150.3 & NIRSPEC & AO     & GL569A \\
2004 Dec 24  & 53363.7 & NIRC2   & AO     & GL569Bab \\
2004 Dec 26  & 53365.7 & NIRSPEC & non-AO & GL569B \\
2005 Feb 22  & 53423.5 & NIRSPEC & non-AO & GL569B  \\
2005 Feb 25  & 53426.5 & NIRC2   & AO     & GL569A,GL569Bab \\
\enddata
\end{deluxetable}

\begin{deluxetable}{lclr}
\tablecaption{Revised Template Library Radial Velocities}
\tablewidth{0pt}
\tablecolumns{4}
\tablehead{
  \colhead{Star} &
  \colhead{Sp. Type} &
  \colhead{Obs. Date} &
  \colhead{V$_{rad}$ (km~s$^{-1}$)}}
\startdata
GL15A   & M1.5 & 2000 Jun 10 & 11.81 \\
GL436   & M2.5 & 2001 Jan 5  & 9.61 \\
GL752A  & M3\phd\phn   & 2000 Jun 9  & 35.88 \\
GL213   & M4\phd\phn   & 2000 Jan 8  & 105.96 \\
GL402   & M4\phd\phn   & 2001 Jan 5  & -1.04 \\
GL669B  & M4.5 & 2000 Jun 10 & -34.90 \\
GL406   & M5.5 & 2001 Jan 5  & 18.61 \\
LHS292\tablenotemark{a} & M6.5 & 2001 Jan 5  & 1.35 \\
GL644C  & M7\phd\phn   & 2001 Feb 2 & 15.21 \\
LHS2351 & M7\phd\phn   & 2001 Feb 2 & 0.72 \\
LHS2065\tablenotemark{b} & M9\phd\phn  & 2001 Feb 2 & 7.02 \\
2MASS 0208+25 & L1\phd\phn & 2002 Dec 15 & 20.79 \\
\enddata
\tablenotetext{a}{Reported to be a spectroscopic binary; $\mathrm{^b}$X-ray flare star (Guenther \& Wuchterl 2003)}
\end{deluxetable}

\begin{deluxetable}{lccl}
\tablecaption{Astrometry of GL~569~AB\label{table_ab}}
\tablewidth{0pt}
\tablecolumns{4}
\tablehead{
  \colhead{Date} &
  \colhead{R.A. $\times$ cos(Dec) (\arcsec)} &
  \colhead{Dec (\arcsec)} &
  \colhead{Reference}}
\startdata
1985.6 & $1.52\pm0.15$ & $4.84\pm0.15$ & Forrest, Skrutskie, \& Shure (1988) \\
1999.7 & $2.11\pm0.08$ & $4.53\pm0.04$ & Mart\'in et al. (2000) \\
2001.1 & $2.45\pm0.22$ & $4.23\pm0.13$ & Lane et al. (2001) \\
2005.2 & $2.72\pm0.14$ & $4.18\pm0.21$ & this work \\
\enddata
\end{deluxetable}

\begin{deluxetable}{lc}
\tablecaption{Radial Velocity Measurements of GL~569~A\label{t_vel_569a}}
\tablewidth{0pt}
\tablecolumns{2}
\tablehead{
  \colhead{MJD} &
  \colhead{$\mathrm{V_{rad}}$ (km~s$^{-1}$)}}
\startdata
52749.4 & -7.1$\pm$1.0 \\
52804.3 & -8.1$\pm$1.0 \\
53150.3 & -7.8$\pm$1.0 \\
\tableline
Average &  -7.7$\pm$0.6 \\
\enddata
\end{deluxetable}

\begin{deluxetable}{lccccccccc}
\tablecaption{Photometry and Astrometry of GL~569~B\label{table_nirc2}}
\tablewidth{0pt}
\tabletypesize{\scriptsize} 
\tablecolumns{10}
\setlength{\tabcolsep}{0.04in}
\tablehead{
  \colhead{MJD} &
  \colhead{Exp. (s)} &
  \colhead{$\mathrm{H_{Ba}}$} &
  \colhead{$\mathrm{H_{Bb}}$} &
  \colhead{$\mathrm{K_{Ba}}$} &
  \colhead{$\mathrm{K_{Bb}}$} &
  \colhead{$\mathrm{\Delta{}m_H}$} &
  \colhead{$\mathrm{\Delta{}m_K}$} &
  \colhead{Sep. (\arcsec)} &
  \colhead{P. A. ($^{\circ}$)}}
\startdata
53363.7 & 100   &\nodata&\nodata&\nodata&\nodata&$0.61\pm0.03$&\nodata&$0.0885\pm0.0048$&$111.3\pm1.2$\\
53426.5 &\phn22 &$10.43\pm0.04$&$11.04\pm0.05$&$9.86\pm0.05$&$10.39\pm0.06$&$0.57\pm0.04$&$0.61\pm 0.04$&$0.0798\pm0.0029$&$133.7\pm0.7$\\
\enddata
\end{deluxetable}

\begin{deluxetable}{lccc}
\tablecaption{Orbital Parameters of GL~569~Bab}
\tablewidth{0pt}
\tablecolumns{3}
\tablehead{
  \colhead{Parameter} &
  \multicolumn{2}{c}{Visual Binary Orbit} \\
 
  \colhead{} &
  \colhead{Zapatero Osorio et al. (2004)} &
   \colhead{This Work} }
\startdata
P (days)       & $876\pm9$     & 863.7$\pm$4.2\\
T (MJD, days)  & $51,822\pm3$  & 51820.9$\pm$2.6\\
$e$            & $0.32\pm0.01$ & 0.312$\pm$0.007\\
$a$ (mas)      &               & 90.4$\pm$0.7\\
$a$ (AU)       & $0.90\pm0.01$\tablenotemark{a} &$ 0.89\pm0.02$\tablenotemark{b}\\
$i$ (deg)      & $34\pm2$      & 32.4$\pm$1.3\\
$\omega$ (deg) & $257\pm2$     & 256.7$\pm$1.7\\
$\Omega$ (deg) & $321.5\pm2.0$ & 321.3$\pm$2.2\\
$M_{total}$ (M$_{\odot}$) & $0.125\pm0.005$\tablenotemark{a} & $0.125\pm0.007$\tablenotemark{b}\\
\enddata
\tablenotetext{a}{Does not include $1.6\%$ parallax uncertainty; $\mathrm{^b}$does include $1.6\%$ parallax uncertainty}
\end{deluxetable}
\clearpage
\setlength{\textheight}{9in}
\begin{deluxetable}{lccccc}
\tablecaption{Radial Velocity Measurements of GL~569~Bab\label{t_vel_569b}}
\tablewidth{0pt}
\tablecolumns{6}
\tablehead{
  \colhead{MJD} &
  \colhead{phase} &
  \multicolumn{4}{c}{V$_{rad}$ (km~s$^{-1}$)} \\

  \colhead{} &
  \colhead{} & 
  \multicolumn{2}{c}{GL~569~Ba} &
  \multicolumn{2}{c}{GL~569~Bb} \\

  \colhead{} &
  \colhead{} &
  \colhead{Order 49} &
  \colhead{Order 48} &
  \colhead{Order 49} &
  \colhead{Order 48}\\

   \hline \\

  \colhead{} &
  \colhead{} & 
  \multicolumn{2}{c}{$~~~~~~$GL644C Templates}}

\startdata

52749.4 & 0.072 & \multicolumn{2}{c}{\phn-7.92$\pm$0.90} & \multicolumn{2}{c}{\phn-9.98$\pm$0.90}  \\
53149.3 & 0.535 & \multicolumn{2}{c}{\phn-8.50$\pm$0.90} & \multicolumn{2}{c}{\phn-8.38$\pm$0.90}  \\
\cutinhead{April 20, 2003 Templates}
52473.3 & 0.753 & -10.89$\pm$1.80    & -11.60$\pm$1.80    & \phn-4.71$\pm$1.40   & \phn-3.19$\pm$1.40 \\
52630.7 & 0.935 & -11.60$\pm$1.80    & \nodata            & \phn-3.65$\pm$1.40   & \nodata            \\
52723.5 & 0.042 & \phn-9.46$\pm$1.80 & \phn-7.12$\pm$1.80 & \phn-8.07$\pm$1.40   & \phn-7.74$\pm$1.40 \\
52804.2 & 0.136 & \phn-7.21$\pm$1.80 & \nodata            & -14.15$\pm$1.40      & \nodata            \\
53030.6 & 0.397 & \phn-7.52$\pm$1.80 & \phn-4.92$\pm$1.80 & -12.72$\pm$1.40      & -11.74$\pm$1.40    \\
53124.4 & 0.506 & \phn-9.07$\pm$1.80 & \phn-8.54$\pm$1.80 & \phn-9.56$\pm$1.40    & \phn-7.29$\pm$1.40 \\
53365.7 & 0.785 & -10.62$\pm$1.80    & -12.31$\pm$1.80    & \phn-4.98$\pm$1.40   & \phn-3.09$\pm$1.40 \\
53423.5 & 0.852 & -10.75$\pm$1.80    & -11.71$\pm$1.80    & \phn-3.99$\pm$1.40   & \phn-1.98$\pm$1.40 \\
\cutinhead{May 24, 2004 Templates}
52473.3 & 0.753 & -12.30$\pm$1.80    & -10.93$\pm$1.80    & \phn-4.57$\pm$1.40   & \phn-3.38$\pm$1.40 \\
52630.7 & 0.935 & -12.12$\pm$1.80    & \nodata            & \phn-3.96$\pm$1.40   & \nodata            \\
52723.5 & 0.042 & \phn-6.92$\pm$1.80 & \phn-4.54$\pm$1.80 & -10.28$\pm$1.40      & \phn-9.98$\pm$1.40 \\
52804.2 & 0.136 & \phn-6.29$\pm$1.80 & \nodata            & -13.21$\pm$1.40      & \nodata            \\
53030.6 & 0.397 & \phn-5.42$\pm$1.80 & \phn-4.49$\pm$1.80 & -13.03$\pm$1.40      & -11.28$\pm$1.40    \\
53124.4 & 0.506 & \phn-6.96$\pm$1.80 & \phn-5.34$\pm$1.80 & -11.64$\pm$1.40      & -11.14$\pm$1.40    \\
53365.7 & 0.785 & -12.90$\pm$1.80    & -12.66$\pm$1.80    & \phn-4.40$\pm$1.40   & \phn-3.75$\pm$1.40 \\
53423.5 & 0.852 & -12.76$\pm$1.80    & -11.82$\pm$1.80    & \phn-3.41$\pm$1.40   & \phn-2.70$\pm$1.40 \\
\enddata
\end{deluxetable}
\clearpage
\setlength{\textheight}{8.4in}
\begin{figure}
\epsscale{1.0}
\plotone{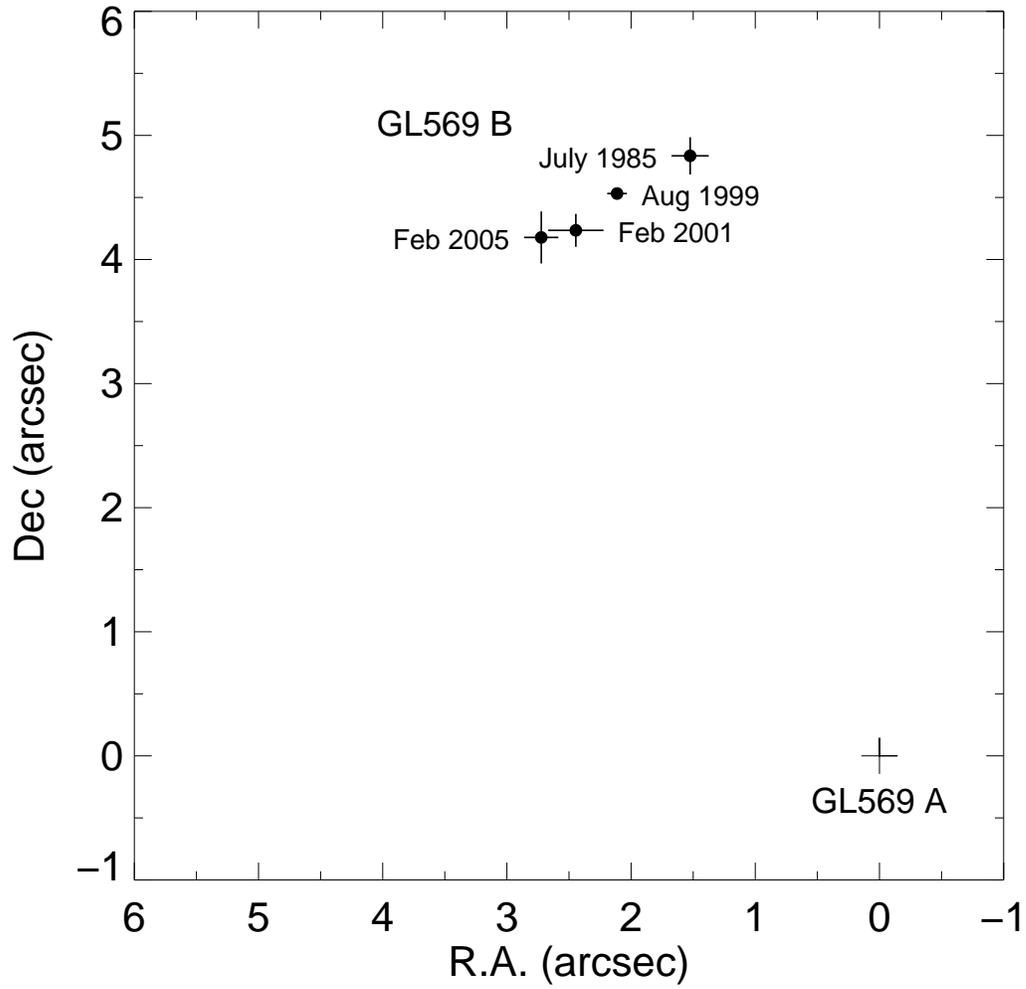}
\caption{The position of GL~569~B with respect to A using our
measurement and the earlier measurements listed in Table 3.  A and B
form a common proper motion system (see \S 3); the figure shows
orbital motion of B with respect to A.  North is up and east is to the
left.}
\label{fig:AB_orb}
\end{figure}

\begin{figure}
\epsscale{1.0}
\plotone{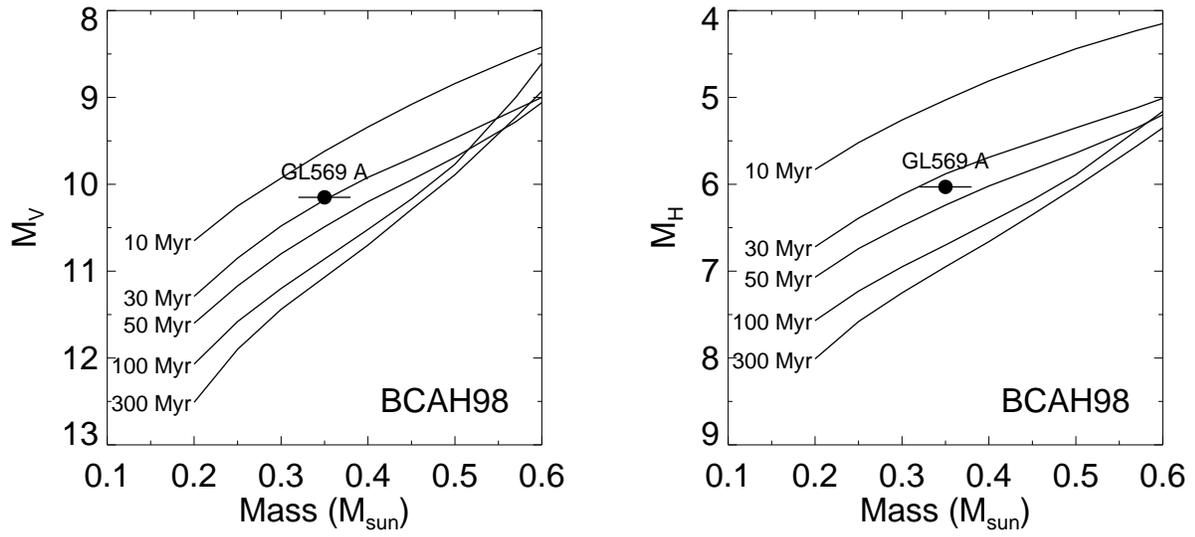}
\caption{Absolute $V$ and $H$ magnitudes and mass of GL~569~A compared
to theoretical isochrones calculated by \citet{bar98}.}
\label{fig:CMD_A}
\end{figure}

\begin{figure}
\epsscale{1.0}
\plotone{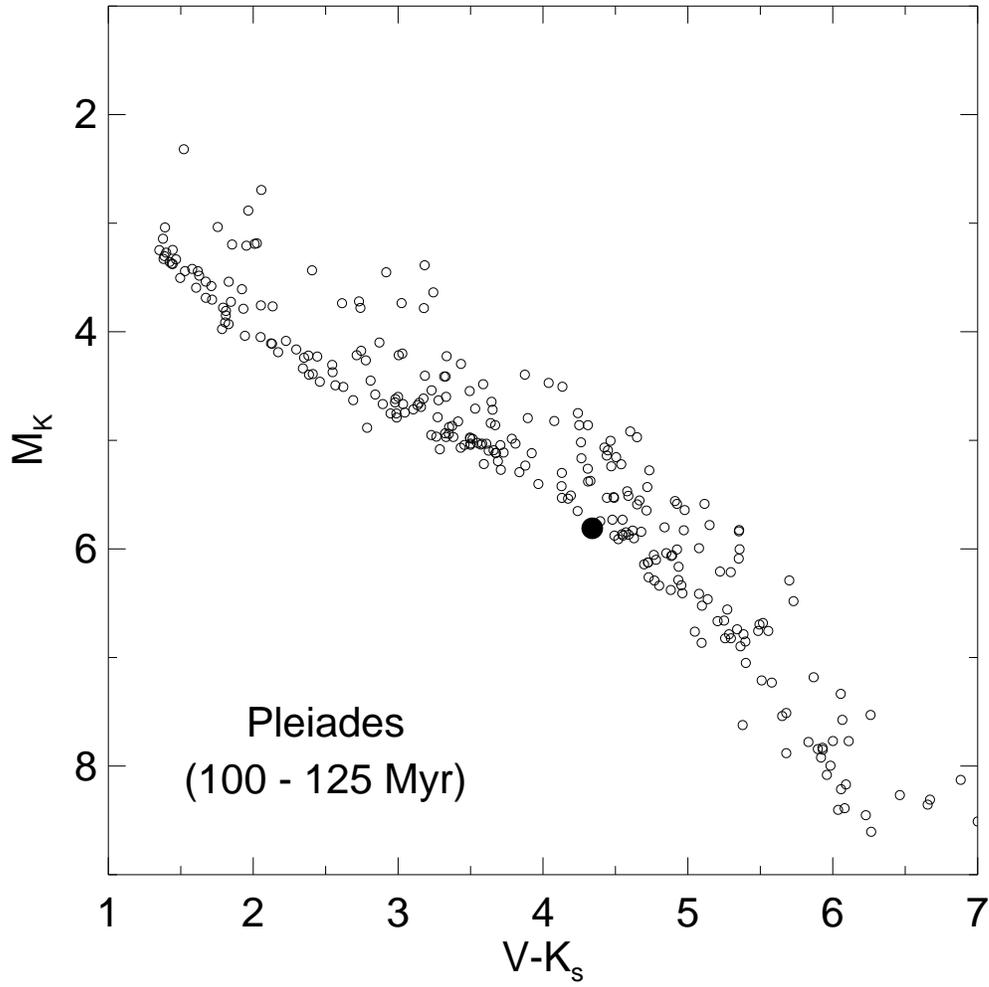}
\caption{Absolute $K$ mag and V$-$K color of GL~569~A (filled circle)
on a color-magnitude diagram of the Pleiades.   The Pleiades data is
the same as in \citet{luh05} Figure 1, and is for an adopted
extinction $A_V=0.12$~mag and distance 133~pc for the Pleiades.}
\label{fig:pleiades}
\end{figure}

\begin{figure}
\epsscale{1.0}
\plotone{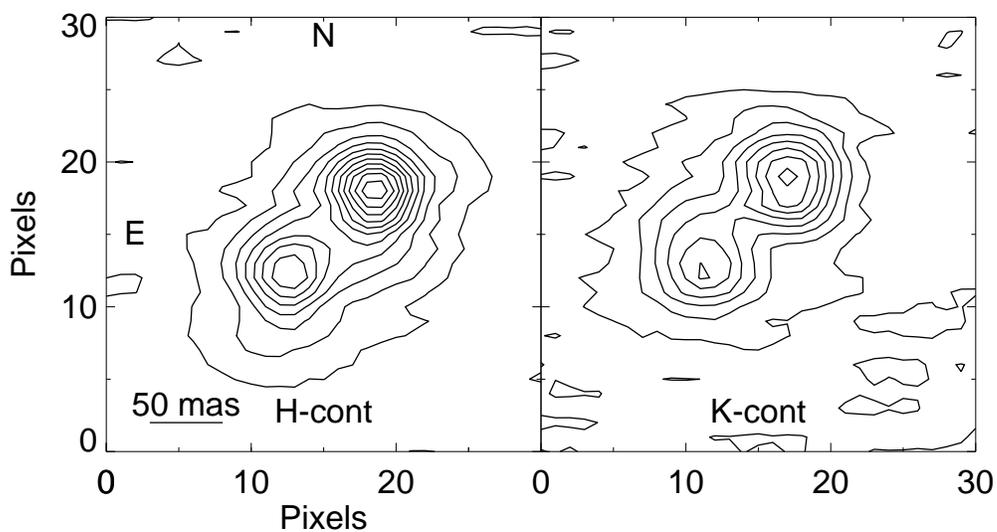}
\caption{NIRC-2 adaptive optics images of GL~569~Ba and Bb at $H$ and $K$
obtained on UT 25 February 2005. North is up and east is
to the left.  The images show a subsection of the frame; GL569~A was
imaged in the same field, but is not shown.  The first contour
represents 1000 counts, $\sim5\sigma$ over the background noise, and
each subsequent contour is an additional 1000 counts.  Ba is located
northwest of Bb and is clearly the brighter component; the Bb/Ba flux
ratios are $0.57\pm0.04$ and $0.61\pm0.04$ in $H$ and $K$
respectively.}
\label{fig:BaBb_imgs}
\end{figure}

\begin{figure}
\epsscale{1.0}
\plotone{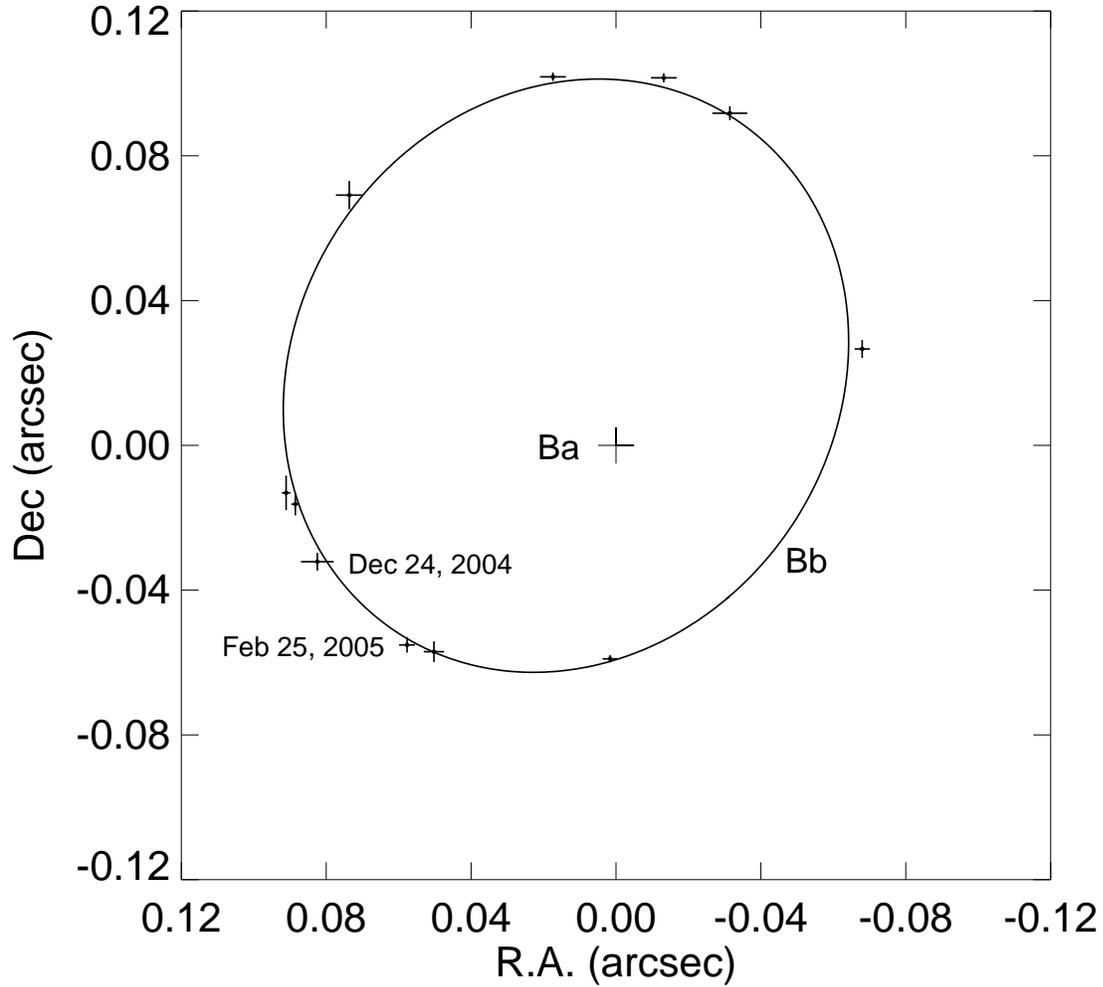}
\caption{The astrometric orbit of GL~569~Bb with respect to Ba at the
coordinate origin, using our measurements from UT 24 December 2004 and
UT 25 February 2005 together with the earlier measurements given in
\citet{zap04}.  The best fit orbit corresponds to our updated
parameters in Table~6.}
\label{fig:BaBb_orb}
\end{figure}
\begin{figure}
\epsscale{1.0}

\plotone{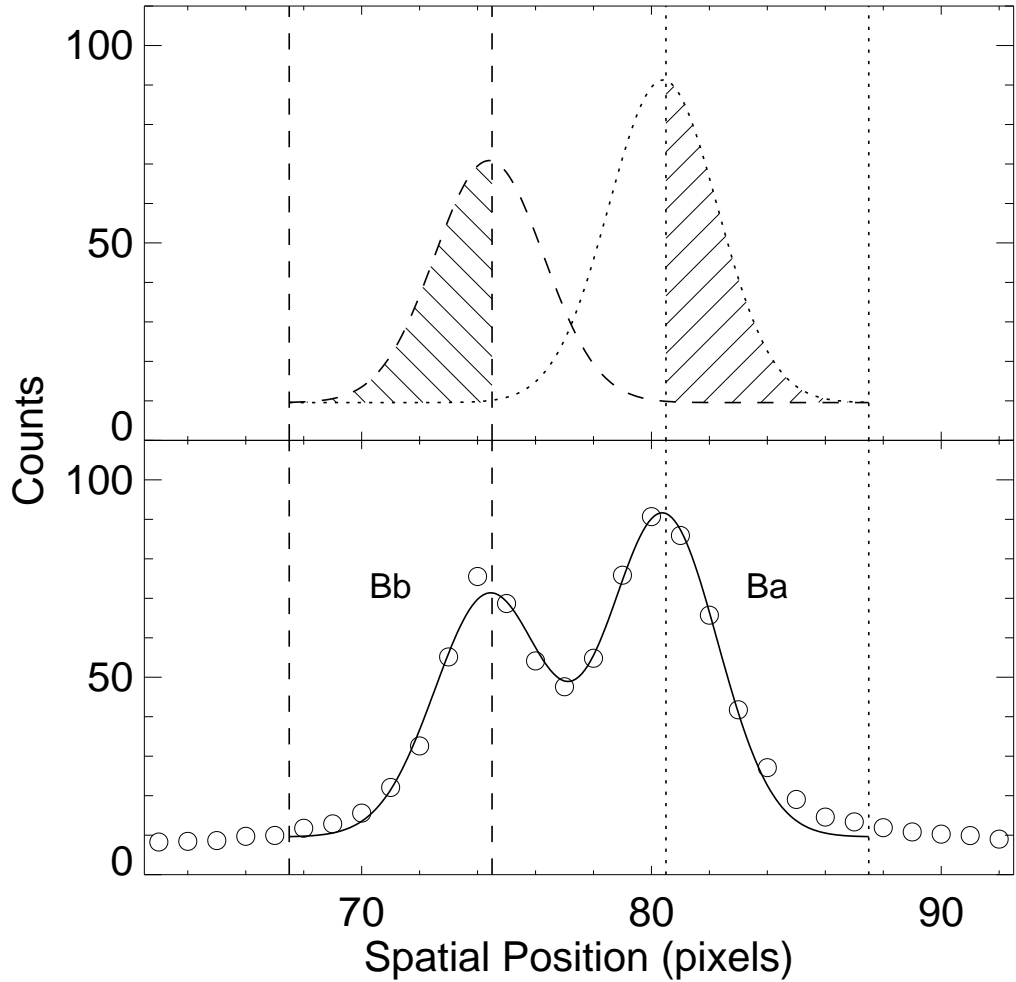}
\caption{A sample profile in the cross-dispersion direction of the AO angularly
resolved Order 49 GL569 Ba and Bb two-dimensional spectrum from UT 24 
May 2004.  The
open circles (bottom panel) show the measured spatial profile of the
resolved spectra; the solid line is a least squares fit of two
Gaussians with equal widths (see text).  The dotted and dashed curves 
(top panel) are the individual Gaussian profiles.  The vertical lines and 
hashed areas indicate the spatial regions used to extract each spectrum.}
\label{fig:BaBb_cut}
\end{figure}

\begin{figure}
\epsscale{1.0}
\plotone{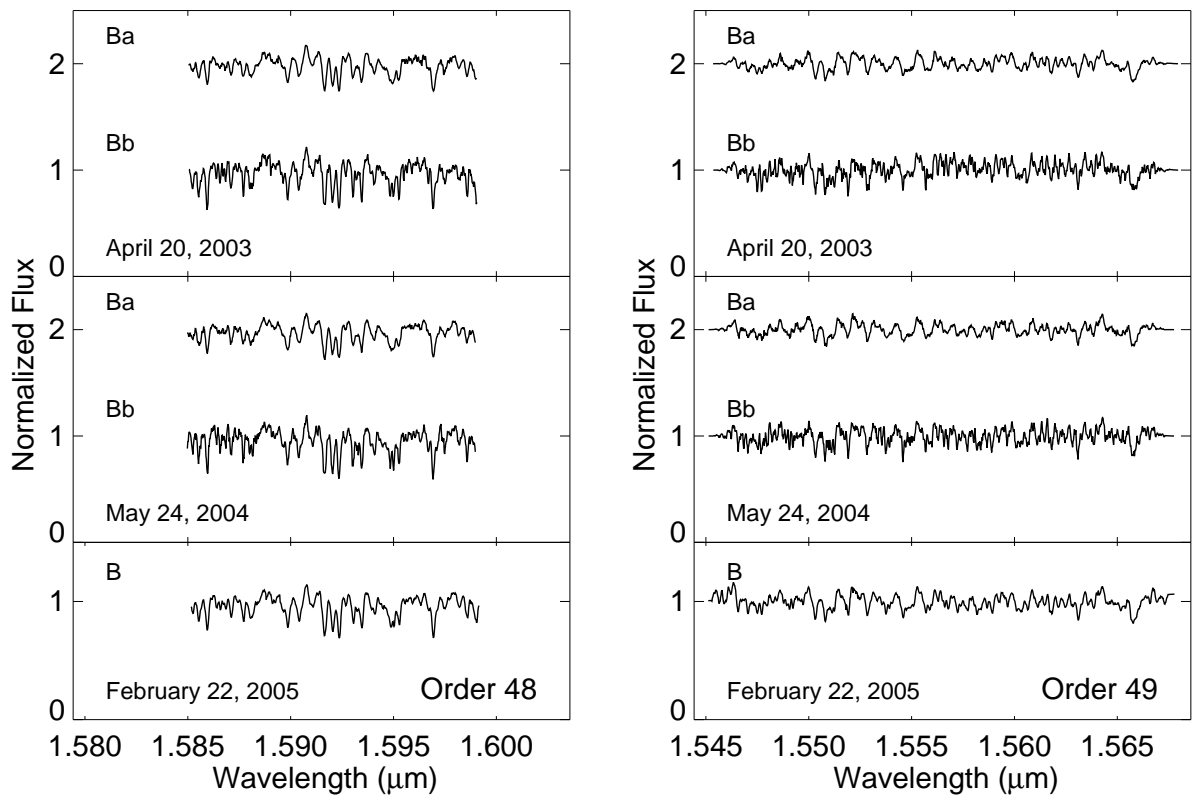}
\caption{Angularly resolved spectra of GL~569~Ba and Bb measured on UT
20 April 2003 and UT 24 May 2004, compared with a spectrum measured on
UT 22 February 2005 in which Ba and Bb are unresolved.  NIRSPEC
spectral orders 48 (left) and 49 (right) are shown.  The wavelength
span of the order 48 spectra is about 2/3 that of the order 49 spectra
because the $\sim$0.007$\mathrm{\,\mu{}m}$ wide section shortward of
$1.585\mathrm{\,\mu{}m}$ is contaminated by a terrestrial CO$_2$
band and not shown. The spectra are normalized to 1 in the continuum 
and are plotted on a 0 to 1 flux scale with the Ba spectra in the top 
two panels offset by +1 in flux.}
\label{fig:targ_sp}
\end{figure}

\begin{figure}
\epsscale{1.0}
\plotone{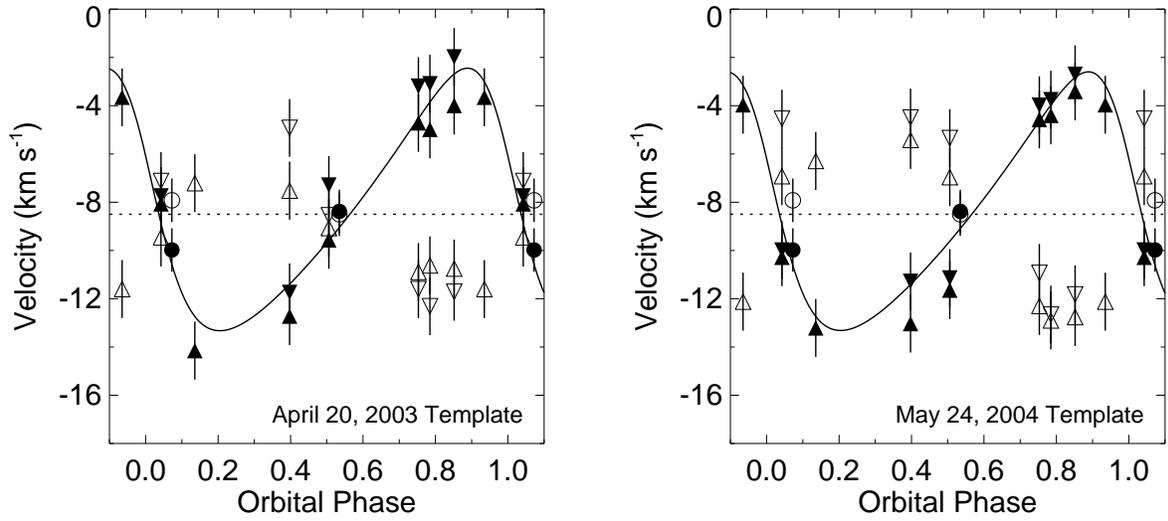}
\caption{Velocity {\it vs} orbital phase measurements for GL~569~Ba
(open triangles) and Bb (filled triangles) using the UT 20 April 2003
templates (left) and UT 24 May 2004 templates (right).  Order 48 and
49 data are plotted as downward and upward pointing triangles,
respectively.  The measurements obtained with AO, averaged over orders
48 and 49, are indicated with open (Ba) and filled (Bb) circles. The
dashed line is the fitted center-of-mass velocity, $-8.5$~km~s$^{-1}$,
and the solid line is the model fit with $K_{Bb}=5.3$~km~s$^{-1}$ and
$P, e, i,$ and $\omega$ given in Table~6.}
\label{fig:vels_ph}
\end{figure}

\begin{figure}
\epsscale{1.0}
\plotone{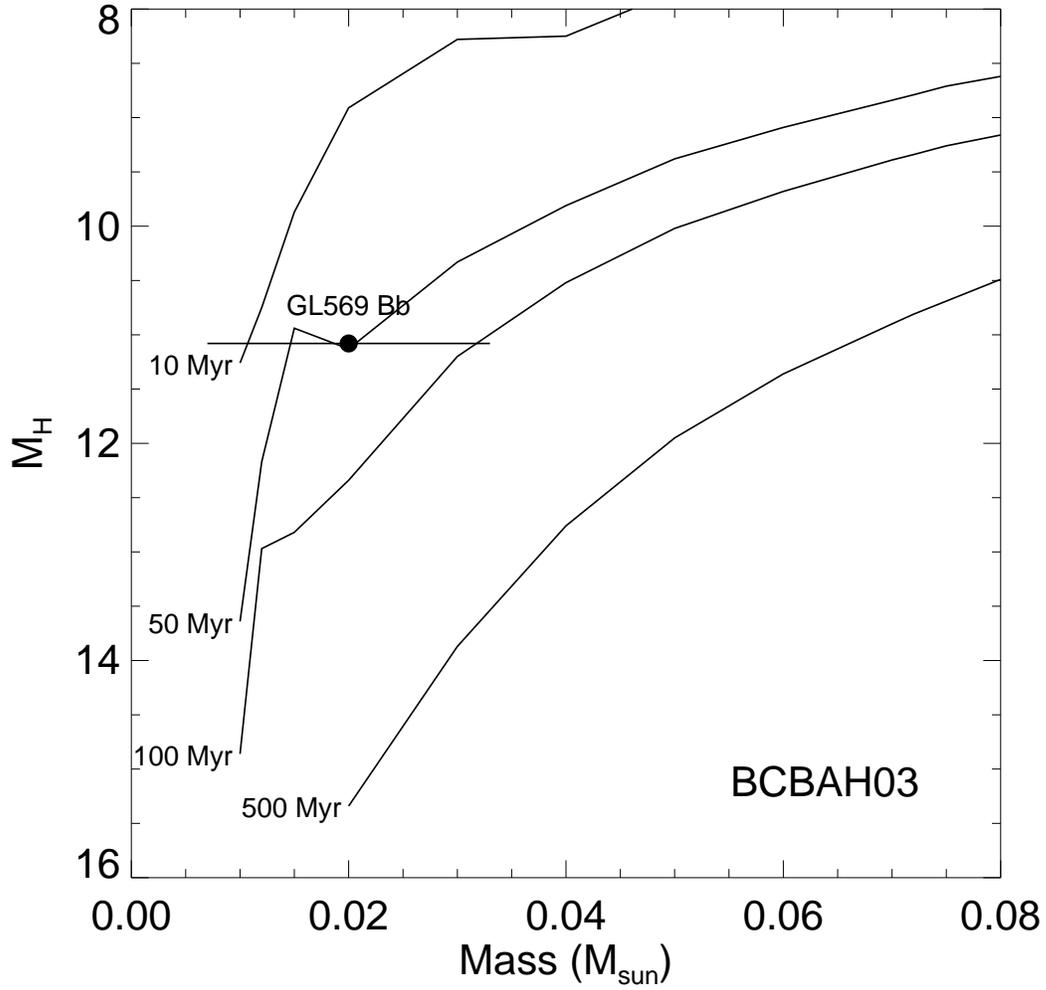}
\caption{GL~569~Bb with M$_H=11.08\pm0.05$ mag and mass
$0.020\pm0.013$~M$_{\odot}$ compared with theoretical isochrones
calculated by \citet{bar03}.  Its location is consistent with an age
of $\sim$100~Myr.}
\label{fig:CMD_Bb}
\end{figure}

\begin{figure}
\epsscale{1.0}
\plotone{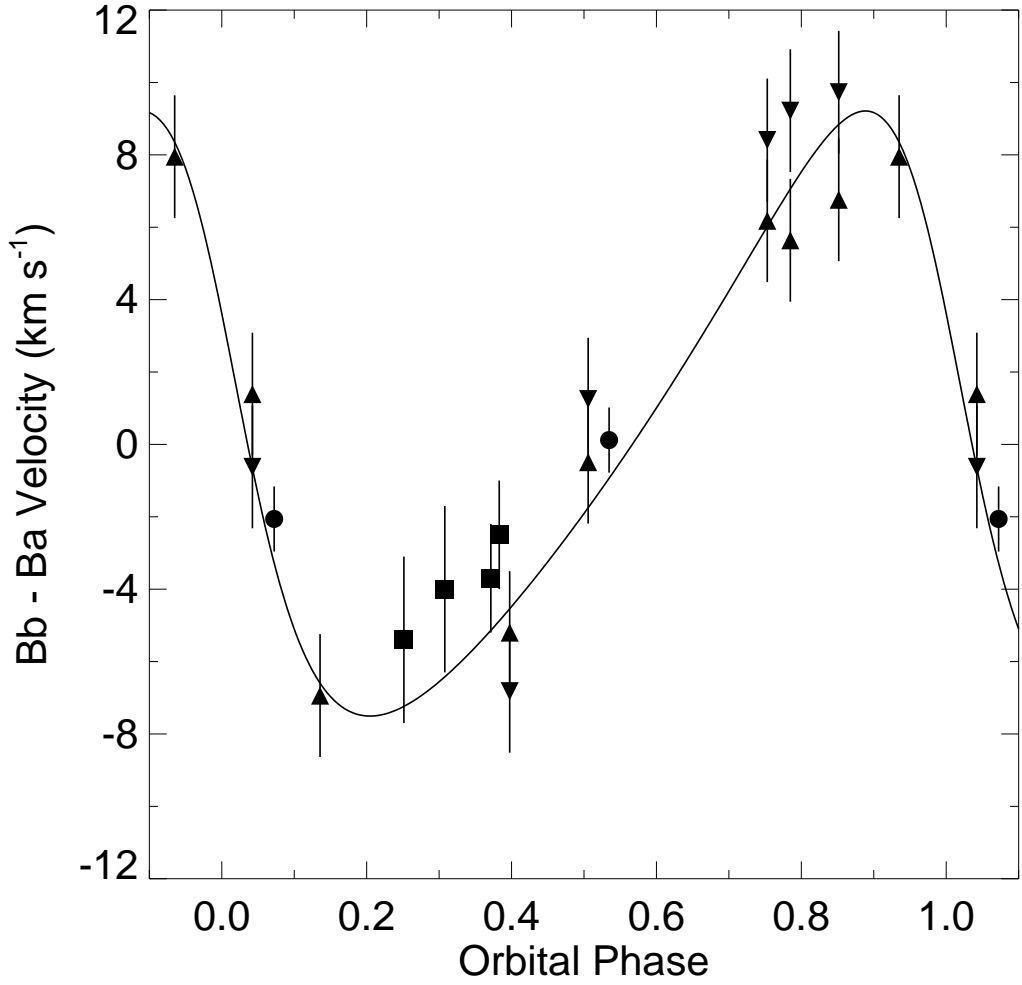}
\caption{Differences of the GL~569~Ba and Bb velocities ${\it vs}$
orbital phase using our measurements in Table~7 with the UT 20 April
2003 templates for: order 48 (downward triangles), order 49 (upward
triangles), and AO (circles).  Our data are consistent with the values
(squares) reported by \citet{zap04}.}
\label{fig:vel_diffs}
\end{figure}

\end{document}